\documentclass[format=acmsmall, review=false]{acmart}
\usepackage{acm-ec-25}
\usepackage{booktabs} 
\usepackage[ruled]{algorithm2e} 

\SetAlFnt{\small}
\SetAlCapFnt{\small}
\SetAlCapNameFnt{\small}
\SetAlCapHSkip{0pt}
\IncMargin{-\parindent}

\usepackage{threeparttable}
\usepackage{tikz}
\usetikzlibrary{shapes}
\usetikzlibrary{calc}
\usetikzlibrary{decorations.pathreplacing}

\setcitestyle{authoryear}

\title[Menu-Based Combinatorial Auctions]{\name: Deep Menus for Combinatorial Auctions by Diffusion-Based Optimization}

\author{Tonghan Wang}\email{twang1@g.harvard.edu}

\affiliation{
  \institution{Harvard University}
  \city{Cambridge}
  \state{MA}
  \postcode{02138}
  \country{USA}
}

\author{Yanchen Jiang}\email{}

\affiliation{
  \institution{Harvard University}
  \city{Cambridge}
  \state{MA}
  \postcode{02138}
  \country{USA}
}

\author{David C. Parkes}\email{}

\affiliation{
  \institution{Harvard University}
  \city{Cambridge}
  \state{MA}
  \postcode{02138}
  \country{USA}
}

\begin{abstract}
Differentiable economics---the use of deep learning for auction design---has driven progress in the automated design of multi-item auctions with additive or unit-demand valuations. However, little progress has been made for optimal combinatorial auctions (CAs), even for the single bidder case, because we need to overcome the challenge of the bundle space growing exponentially with the number of items. For example, when learning a menu of allocation-price choices for a bidder in a CA, each menu element needs to efficiently and flexibly specify a probability distribution on bundles. In this paper, we solve this problem in the single-bidder CA setting by generating a bundle distribution through an ordinary differential equation (ODE) applied to a tractable initial distribution, drawing inspiration from generative models, especially score-based diffusion models and continuous normalizing flow. Our method, \name, uses deep learning to find suitable ODE-based transforms of initial distributions, one transform for each menu element, so that the overall menu achieves high expected revenue.  Our method achieves 1.11$-$2.23$\times$ higher revenue compared with automated mechanism design baselines on the single-bidder version of CATS, a standard CA testbed, and scales to problems with up to 150 items. Relative to a baseline that also learns allocations in menu elements, our method reduces the training iterations by 3.6$-$9.5$\times$ and cuts training time by about 80\% in settings with 50 and 100 items.
\end{abstract}

\usepackage{amsmath}
\usepackage{mathtools}
\usepackage{amsthm}
\usepackage{wrapfig}

\usepackage[capitalize,noabbrev]{cleveref}

\makeatletter
\def\thmheadbrackets#1#2#3{%
  \thmname{#1}\thmnumber{\@ifnotempty{#1}{ }\@upn{#2}}%
  \thmnote{ {\the\thm@notefont[#3]}}}
\makeatother

\newtheoremstyle{brakets}
  {}
  {}
  {\itshape}
  {}
  {\bfseries}
  {.}
  { }
  {\thmheadbrackets{#1}{#2}{#3}}

\theoremstyle{brakets}

\theoremstyle{remark}

\definecolor{darkgreen}{rgb}{0.0, 0.5, 0.0}
\definecolor{darkblue}{rgb}{0.0, 0.5, 1.0}

\newcount\Comments  
\newcommand{\kibitz}[2]{\ifnum\Comments=1{\color{#1}{#2}}\fi}
\newcommand{\tw}[1]{\kibitzAdd{tw}{[Tonghan: #1]}}
\newcommand{\dcp}[1]{\kibitz{teal}{[DCP: #1]}}

\newcount\CommentsAdd  
\newcommand{\kibitzAdd}[2]{\ifnum\CommentsAdd=1{\color{#1}{#2}}\fi}
\definecolor{english}{rgb}{0.0, 0.5, 0.0}
\definecolor{tw}{rgb}{0.0, 0.0, 0.5}

\usepackage{xcolor}

\usepackage{amsmath,amsfonts,bm}

\def\eqref#1{equation~\ref{#1}}

\def\1{\bm{1}}

\def\vb{{\bm{b}}}

\def\vf{{\bm{f}}}

\def\vs{{\bm{s}}}

\def\vw{{\bm{w}}}
\def\vx{{\bm{x}}}

\def\mI{{\bm{I}}}

\DeclareMathAlphabet{\mathsfit}{\encodingdefault}{\sfdefault}{m}{sl}
\SetMathAlphabet{\mathsfit}{bold}{\encodingdefault}{\sfdefault}{bx}{n}

\DeclareMathOperator{\Tr}{Tr}

\newcommand{\shortn}{\textup{\texttt{-}}}

\newcommand{\ie}{\textit{i}.\textit{e}.}
\newcommand{\eg}{\textit{e}.\textit{g}.}

\usepackage{multirow}
\usepackage{makecell}
\newcolumntype{L}{>{$}l<{$}}
\newcolumntype{C}{>{$}c<{$}}
\newcolumntype{R}{>{$}r<{$}}

\newcommand{\name}{\textsc{BundleFlow}}
\newcommand{\bundle}{\texttt{Bundle-RochetNet}}
\newcommand{\bigbundle}{\texttt{Big-Bundle}}
\newcommand{\smallbundle}{\texttt{Small-Bundle}}
\newcommand{\grandbundle}{\texttt{Grand-Bundle}}

\begin{document}

\begin{titlepage}

\maketitle\makeatletter \gdef\@ACM@checkaffil{} \makeatother
\setcounter{tocdepth}{2} 

\end{titlepage}

\section{Introduction}


When selling multiple items simultaneously, bidders may have complex valuations that exhibit synergies among items. For instance, some items may act as complements, making their collective value to a bidder exceed the sum of their individual values. \emph{Combinatorial auctions} (CAs) support these kinds of valuations by allowing bids on bundles of items. The need for such auctions was recognized as early as 1922~\cite{uscongress1925e}, and their formal definition dates to 1982~\cite{rassenti1982combinatorial}, where they are exemplified in the allocation of congested airport runways. Since then, CAs have proven pivotal in addressing a wide range of real-world challenges, most notably in the auctioning of spectrum licenses~\cite{cramton1997fcc,palacios-huerta24}--efforts whose far-reaching impact contributed to the awarding of the 2020 Nobel Prize in Economic Sciences~\cite{NobelPrizeEcon2020}.

Despite their  prominence, designing optimal CAs remains fundamentally challenging. As with auctions for additive or unit-demand valuations, it is typical to seek mechanisms that are (1) dominant-strategy incentive compatible (DSIC, also strategy-proof), ensuring that bidders benefit most by reporting their true values, and (2) revenue-maximizing from the auctioneer's perspective. However, even the seemingly simpler single-bidder combinatorial setting---a foundational building block for multi-bidder scenarios---still lacks a comprehensive theoretical characterization. This gap highlights the broader difficulty in developing and analyzing optimal mechanisms for general CAs.

In response to similar theoretical obstacles in additive and unit-demand valuations, researchers have explored deep learning techniques, commonly referred to as \emph{differentiable economics}~\cite{dutting2024optimal}. In particular, deep menu-based methods show promise. These methods learn a menu of options for a bidder, guaranteeing strategy-proofness, provided the menu remains self-bid independent and agent-optimizing~\citep{hammond1979straightforward}. 
For a CA, each option in a menu will correspond to a bundle of items (or a distribution on bundles) and a price. Following {\em RochetNet}~\cite{dutting2024optimal}, various methods have been developed for the single-bidder but non-combinatorial setting~\cite{curry2022differentiable,duan2023scalable,shen2019automated,dutting2024optimal}, demonstrating
the ability to rediscover auctions that are provably optimal. {\em GemNet}~\cite{wang2024gemnet} extends menu-based methods to multi-bidder scenarios and pushes the frontier for the design of multi-item auctions with additive or unit-demand
valuations. 

Unfortunately, differentiable economics has made only limited headway on the problem of optimal CA design and the fundamental challenge of handling the exponential number of bundles remains largely untouched. \citet{dutting2024optimal} deal with two items, and with a learned mechanism that does not guarantee exact DSIC. \citet{duan2024scalable} scale to 10 items, but restrict their attention to  the virtual valuation combinatorial auction (VVCA), which is not a fully general design space. \citet{ravindranath2024deep} consider a sequential CA setting. None of these studies provide a path towards DSIC and expressive, \ie, fully general, mechanisms for tens or hundreds of items, 
which presents a formidable obstacle and necessitates the development of novel methodology.

Focusing on the single-bidder setting, we make progress by developing a menu-based, deep learning method for DSIC and expressive CAs that scales to as many as 
150 items.   Single-bidder CAs, although less general than multi-bidder CAs, are applicable to  real-world problems. Consider  a digital content provider offering a collection of movies to a viewer, a cloud computing vendor offering different features (clock speed, number of compute units,  background execution, etc.), a monopolist seller offering  highly complementary patents to a  pharmaceutical firm, or a utility  supplier bundling electricity, gas, and renewable energy for an industrial customer. In these examples, the buyer's valuation may be influenced by strong complementarities or substitutes across items.
Moreover, and as discussed in Sec.~\ref{sec:multi-bidder}, 
this single-bidder deep mechanism design algorithm, which extends generative models to solve
the bundle scalability issue, provides a 
direction towards the automated design of multi-bidder CAs at scale.

\begin{figure}
    \centering
    \includegraphics[width=\linewidth]{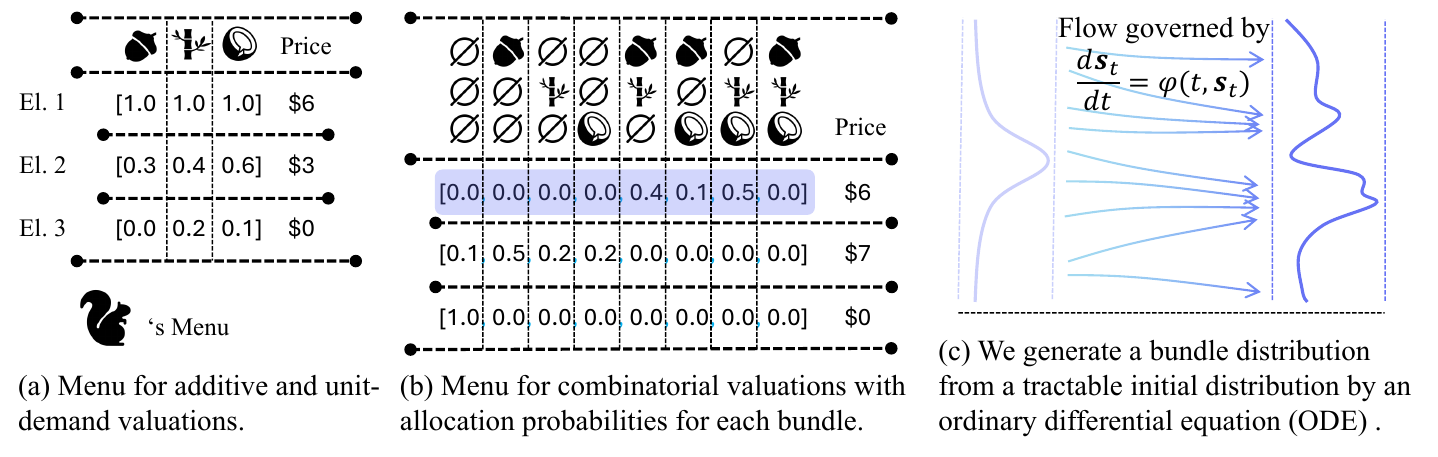}
    \caption{(a) A menu for additive or unit-demand valuations only needs to specify allocation probabilities for each item. However, item-wise allocation probabilities are  too inflexible
    for CAs, as bidder values are specified for bundles. (b) The space complexity for representing an explicit distribution on bundles (a bundle-wise allocation) grows exponentially with the number of items. (c) We represent a bundle distribution
    through a tractable initial distribution and an ordinary differential equation (ODE). 
    \label{fig:idea}}
\end{figure}
To understand the challenge of optimal CA design, even with one bidder, 
Fig.~\ref{fig:idea} compares  menus for additive and unit-demand valuations with menus for combinatorial valuations.
For an additive bidder, it suffices to specify in a menu element (or option) the \emph{item-wise allocation probabilities} such as $[0.3, 0.4, 0.6]$ along with a price. A bidder's  expected value for the corresponding menu element can be calculated as the weighted sum of item values. A similar, item-wise  approach also works for a unit-demand bidder.
However, such item-wise allocations are ambiguous in CAs.
For example, $[0.3, 0.4, 0.6]$ could be $0.3[1,0,0]+0.4[0,1,0]+0.6[0,0,1]$ or
$0.3[1,1,1]+0.1[0,1,1]+0.2[0,0,1]$, with the same item-wise allocation  decomposed as different \emph{bundle-wise allocations}, leading to conflicting valuations. One could predefine how to interpret a marginal item-wise allocation to avoid ambiguity, for example adopting the product distribution semantics. However, a product distribution requires computation that is exponential in the number of items to evaluate the value for a bidder with a general, combinatorial valuation function. More importantly, the use of a product distribution lacks flexibility; e.g.,  $0.5[1,1,0]+0.5[0,0,1]$ cannot be represented as a product distribution. Indeed, any fixed mapping from the $m$-dimensional space ($m$ is the number of items) of item-wise allocations to the $2^m$-dimensional space of bundle-wise allocations means that many bundle-wise allocations are left uncovered. This lack of expressiveness is reflected in our experiments--learning product distributions (\eg, \bundle~in Table~\ref{tab:exp_results}) lags behind some fixed-allocation menus.

One way to achieve expressiveness would be to explicitly specify, for each menu element, an allocation probability for each possible bundle, as shown in Fig.~\ref{fig:idea} (b). However, the number of bundles grows exponentially with the number of items, making it intractable to learn these bundle-wise allocations. Aside from menu-based approaches, other state-of-the-art deep learning methods for mechanism design face a similar challenge: they do not suggest a way to represent bundle-wise allocation distributions efficiently and flexibly.
%

%

In this paper, we solve this problem by avoiding the need to directly represent and learn an exponentially high-dimensional specification of a distribution on bundles. Instead, we represent a distribution on bundles by a tractable and low-dimensional \emph{initial distribution}, $\alpha_0(\vs_0)$, on {\em initial bundle variables}, $\vs_0\in \mathbb{R}^m$,
and an ordinary differential equation (ODE): $d\vs_t = \varphi(t, \vs_t) dt$. The ODE operates on \emph{bundle variables}, $\vs_t\in\mathbb{R}^m$, through the {\em vector field} $\varphi(t, \cdot)$. Here $m$ is the number of items and $t\in[0,T]$ is the {\em ODE time}. A feasible bundle corresponds to a bundle variable where the entries are all 0s or 1s. Formally, $\vs_t$ is generated from a sample $\vs_0$ by applying the ODE: $\vs_t(\vs_0)=\vs_0+\int_0^t \varphi(\tau,\vs_\tau) d\tau$. We omit the dependence on $\vs_0$ and write $\vs_t$ for simplicity.
In this way, the ODE transforms the initial distribution $\alpha_0(\vs_0)$ to a final distribution $\alpha_T(\vs_T)$ at time $T$. The idea is to (1) train the vector field so that the support of the final distribution $\alpha_T(\vs_T)$ corresponds to feasible bundles; (2) fix this trained vector field $\varphi(t, \cdot)$, and learn a different initial distribution $\alpha_0(\vs_0)$ (therefore a different final distribution over bundles) and price for each menu
element, so that the menu maximizes expected revenue. 

\begin{figure}
    \centering
    \includegraphics[width=\linewidth]{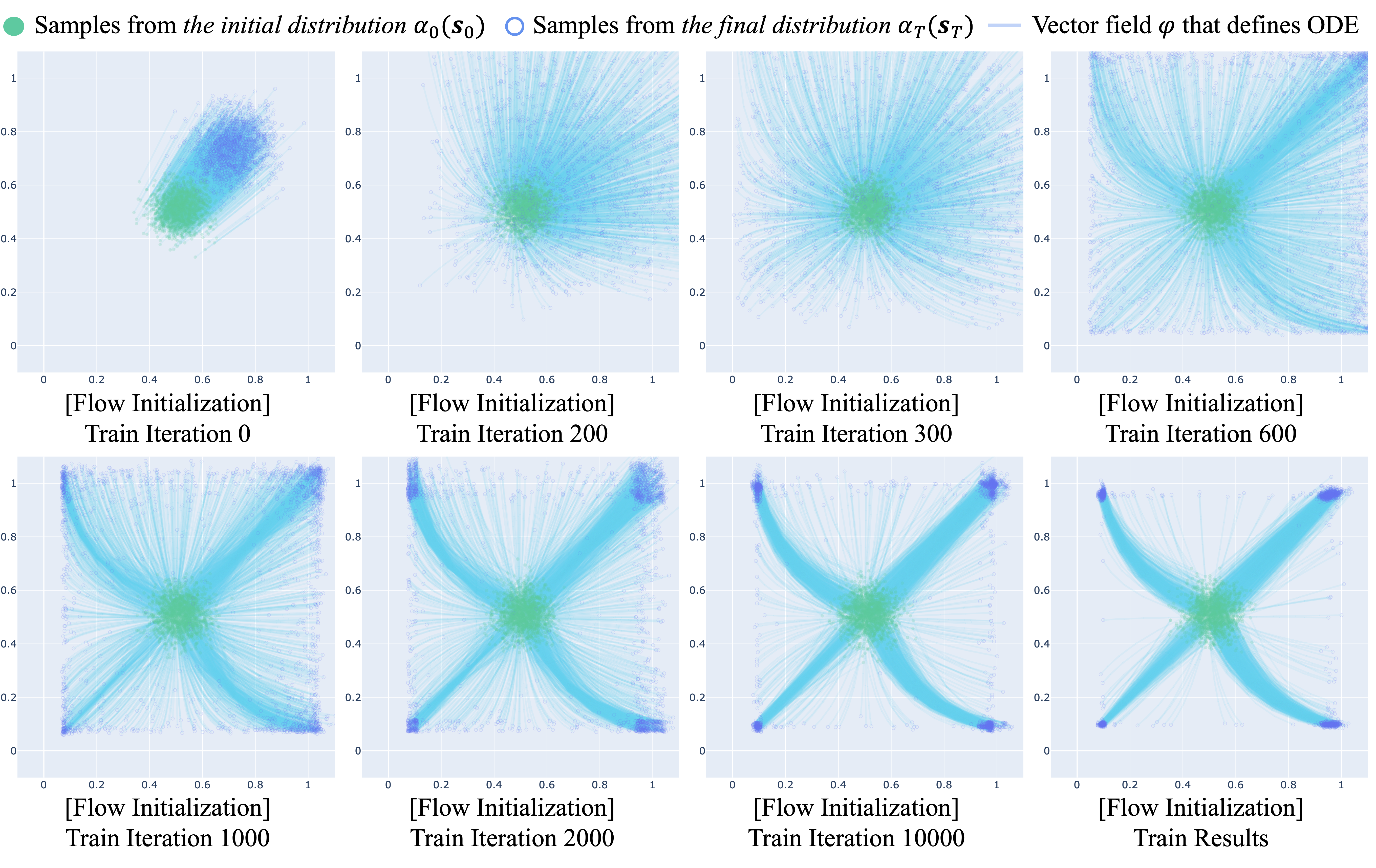}
    \caption{Evolution of the vector field $\varphi$ (represented by blue curves) during the first stage of menu training: \emph{Flow Initialization}. The x- and y-axes represent the bundle variables for two of items. $x=1$ means item A is in the bundle, and $y=1$ means item B is in the bundle. We employ an ODE $d\vs_t = \varphi(t, \vs_t) dt$ to generate the final distribution $\alpha_T(\vs_T)$ (a distribution over bundles), represented by blue dots, from a simple initial distribution $\alpha_0(\vs_0)$, represented by green dots. During the first stage, $\alpha_0(\vs_0)$ is fixed as a mixture-of-Gaussian distribution. Dot opacity represents probability density. The aim of this first stage is to train the vector field so that the final distribution has all feasible bundles as its support (see Sec.~\ref{sec:viz}).
    \label{fig:flow_init}}
\end{figure}
This method draws inspiration from generative AI models, such as {\em diffusion models}~\cite{ho2020denoising, kadkhodaie2023generalization,song2021score,song2019generative,yang2025policy} and, in particular, {\em continuous normalizing flow}~\cite{chen2018neural,liu2022flow,lipman2022flow}.  We thus call the  method \name~to emphasize the core idea of using continuous normalizing flow to model bundle distributions. Prior work has successfully shown how to transform simple distributions such as the Gaussian distribution to complex \emph{target distributions} such as natural images~\cite{rombach2022high,esser2024scaling}, language~\cite{lou2024discrete}, and videos~\cite{stabilityAI2023}. In these generative AI tasks, the target distribution is known and observed as the data distribution in large-scale pre-training datasets. Our technical novelty is to extend generative models to solve optimization problems, seeking a bundle distribution (and price) for
each menu element that optimizes expected revenue. 
There is no known target distribution in our work.


Central to our method is that the distribution's evolution under the ODE
is governed by the {\em Liouville equation}~\cite{liouville1838note}:
\begin{align}
    \log \alpha_t(\vs_t) = \log \alpha_0(\vs_0) -\int_0^t \nabla\cdot\varphi(\tau, \vs_\tau) d\tau.\label{equ:demo_liouville}
\end{align}
We design the functional form of the vector field $\varphi(t, \cdot)$ so that the integral of its divergence $\nabla\cdot\varphi(t, \cdot)$ is easy to compute, thereby allowing efficient menu optimization. The initial distribution $\alpha_0(\vs_0)$ is chosen to be a simple distribution. As the first stage of menu learning, we fix $\alpha_0(\vs_0)$ to a mixture-of-Gaussian distribution 
and train a single vector field $\varphi(t, \cdot)$ to transport any initial variable $\vs_0$ to a feasible bundle at final time $T$. An illustration of how the vector field evolves during this first training stage is shown in Fig.~\ref{fig:flow_init}.

\begin{figure}
    \centering
    \includegraphics[width=\linewidth]{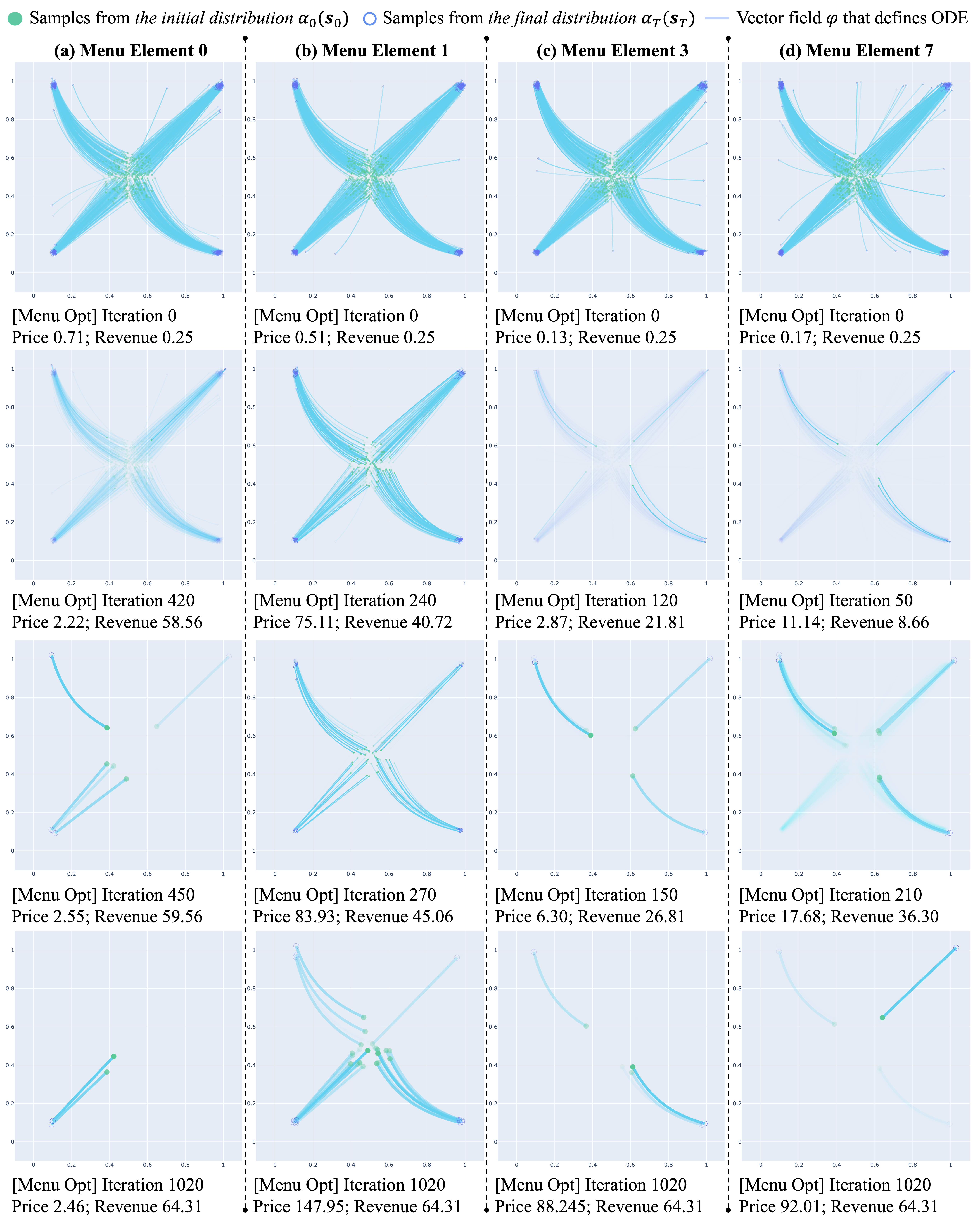}
    \caption{Visualization of the second stage of training: \emph{Menu Optimization}. The figure presents snapshots of four menu elements (organized in columns) at different training iterations (organized in rows), showing the bundle distribution and price for each element, along with the test-time auctioneer revenue from the entire menu at the corresponding iteration. The x- and y-axes represent the bundle variables for two of items. $x=1$ means item $A$ is in the bundle, and $y=1$ means item $B$ is in the bundle. We fix the vector field (blue curves) and update initial distributions of elements to manipulate distributions over bundles (refer to Sec.~\ref{sec:viz}).
    \label{fig:flow_train}}
\end{figure}

With this vector field fixed, as a second stage we then train each element in a menu so that the menu maximizes expected auction revenue.  The advantage of our flow-based method is that, during both training and testing, we can efficiently calculate the utility of each menu element $k$ by substituting the bundle distribution $\alpha_T^{(k)}(\vs_T)$ with $\alpha_0^{(k)}(\vs_0)$ as described in Eq.~\ref{equ:demo_liouville}.
We re-parametrize the initial distribution $\alpha_0^{(k)}(\vs_0)$ so that the resulting bundle distribution, $\alpha_T^{(k)}(\vs_T)$, becomes trainable.
To ensure DSIC, we design the initial distribution $\alpha^{(k)}_0(\vs_0)$ to have finite support. Specifically, we use mixture-of-Dirac distributions.
This enables precise reconstruction of the bundle distribution $\alpha_T^{(k)}(\vs_T)$ and thus the utility of a menu element by enumerating the support of $\alpha_0^{(k)}(\vs_0)$.

For optimization, each menu element shares a common vector field and has separate trainable parameters for its price and  initial distribution (mixture weights and support points of the mixture-of-Dirac distribution).  
The gradients of the revenue-maximizing loss backpropagate through the ODE back to parameters of $\alpha_0^{(k)}(\vs_0)$, guiding updates that increase expected revenue. 
As a demonstration of  this second stage, we illustrate changes of the initial distribution and the corresponding bundle  distribution for each of four different menu elements, as well as corresponding prices and auction revenue in Fig.~\ref{fig:flow_train}. 


Although the number of bundles in the support of the bundle distribution $\alpha^{(k)}_T(\vs_T)$ for menu element $k$ is 
at most the size of the support of the initial distribution $\alpha^{(k)}_0(\vs_0)$, what this representation achieves
is that the bundle distribution $\alpha^{(k)}_T(\vs_T)$ can be flexibly learned.
In practice, we observe that even a small support size for bundle distribution $\alpha^{(k)}_T(\vs_T)$ yields strong performance across various settings. Moreover, the formulation of our method does not rely on any  assumptions regarding the bidder's valuation function, for example that the valuation function can be represented on a small number 
of bundles, and thus exhibits flexibility, applicable to a large range of CA problems.
%
%

We evaluate \name~using single-bidder instantiations of CATS \citep{leyton2000towards}, a widely recognized CA testbed. Given the absence of established deep methods capable of extending to the settings considered in this paper, we benchmark against the following DSIC baselines. (1) \bundle: An adaptation of RochetNet~\cite{dutting2024optimal} with a menu of item-wise allocations (and prices) that interprets item-wise allocations as product distributions. (2) \bigbundle: This baseline fixes allocations in all menu elements, favoring bundles with the highest item counts. Prices are learned using the same gradient-based method as in RochetNet. (3) \smallbundle: Similar to \bigbundle\ but in addition to the grand bundle, which includes all items, it also includes bundles with minimal item allocations. (4) \grandbundle: Employs a grid search to determine a price for the grand bundle. In all cases, the menu element with the largest utility, if non-negative, is assigned to the bidder. 
These baselines represent progressively restrictive constraints on the flexibility of menus.

Experimental results demonstrate that our method consistently and significantly outperforms all baselines across all benchmark settings and scales effectively to auctions involving up to 150 items. Specially, for auctions with 50 to 150 items, \name~achieves 1.11$-$2.23$\times$ higher revenue. This enhanced revenue performance does not compromise training efficiency. On the contrary, compared to the baseline \bundle~that also learns allocations in menu elements, our method typically requires 3.6$-$9.5$\times$ fewer training iterations and can reduce training time by about 80\% in settings with 50 or 100 items.

A critical observation is noted when we vary the support size of initial distributions, $D$, from 2 to 1. This change results in a considerable revenue decline in \name, causing its performance to become comparable to that of the baselines, with a reduction of up to 52.7\% at its most severe. Since when $D=1$ each menu element deterministically assigns a single bundle, these results underscore the importance of allowing randomized distributions over bundles in differentiable economics for CA settings, a capability uniquely enabled by our method.



\section{Preliminaries}

\textbf{Sealed-Bid Combinatorial Auction}. We consider sealed-bid CAs with a single bidder and $m$ items, $M=\{1,\ldots,m\}$.
The bidder has a {\em valuation function}, $v: 2^M \rightarrow \mathbb{R}_{\ge 0}$. Valuation $v$ is drawn independently from a distribution $F$ defined on the space of possible valuation functions $V$, determining how valuable each bundle $S\in 2^M$ is for the bidder. We consider bounded valuation functions: $v(S)\in[0, v_{\max}]$, $S\subset 2^M$, with $v_{\max}>0$, and they are normalized so that $v(\varnothing)=0$.
%
The auctioneer  knows distribution $F$ but not the  valuation  $v$. The bidder reports their valuation function, perhaps untruthfully, as their {\em bid (function)}, $b\in V$. 

In CAs, a suitable {\em bidding language} is critical to allow a bidder to report their
bid without needing to enumerate a value for every possible bundle.  There
are many ways to do this, but a  common approach is to use the {\em XOR bidding language}, which allows bidders to submit bid prices for each of multiple bundles under an exclusive-or condition; in effect, only one bid price on a bundle can be accepted. Popular CA testbeds such as CATS~\citep{leyton2000towards} and SATS~\citep{weiss2017sats} employ this bidding language extensively.\footnote{When representing the values of multiple bidders these testbeds often also introduce so-called dummy items for distinguishing the bids of different 
bidders. Still, the semantics for a single bidder is, in effect, that of the XOR language.}
The semantics of the XOR bidding language is that the value on a bundle $S$ is the maximum bid price on
any bundle $S'$, submitted as part of the XOR bid, and for which $S'\subseteq S$. XOR bids are 
succinct for valuation functions in which the bidder is only interested in a bounded number
of possible bundles.

We seek an auction $(g,p)$ that maximizes expected revenue. Here, $g: V\rightarrow \mathcal{X}$ is the {\em allocation rule},  where $\mathcal{X}$ is the space of feasible allocations (i.e., no item allocated more than once), so that $g(b)\subseteq M$ denotes the set of items (perhaps empty) allocated to the bidder at bid $b$.
Also, $p: V\rightarrow \mathbb{R}_{\ge 0}$ is the {\em payment rule},
specifying the price associated with allocation $g(b)$. 
The utility to the bidder with valuation function $v$ at bid  $b$ is $u(v;b)=v(g(b))-p(b)$, which is
the standard model of quasi-linearity so that values are in effect quantified in monetary units, say dollars.
 In full generality, the allocation and payment rules may be \emph{randomized}, with
 the bidder assumed to be risk neutral  and seeking
to maximize their 
expected utility.

In a \emph{dominant-strategy incentive compatible} (DSIC) auction, or {\em strategy-proof (SP)} auction, the bidder's utility is  maximized by bidding their true valuation $v$, whatever this valuation is; i.e., $u(v; v)\ge u(v;b)$, for $\forall v\in V, \forall b\in V$.
An auction is \emph{individually rational} (IR) if the bidder receives a non-negative utility when participating and truthfully reporting: $u(v;v)\ge 0$, for $\forall v\in V$.
Following the revelation principle, it is without loss of generality to focus on  
SP
auctions, as any auction that achieves a particular expected revenue in a dominant-strategy equilibrium
 can be transformed into an SP auction with the same expected revenue.
 Optimal auction design therefore seeks to identify an SP and IR auction that maximizes the expected revenue, i.e., $\mathbb{E}_{v\sim \bm F}[p(v)]$. 

\textbf{Menu-Based CAs}. In a {\em menu-based auction}, allocation and payment rules  are
represented through a menu, $B$, consisting of
$K\ge 1$  {\em menu elements}.
We write $B=(B^{(1)},\ldots,B^{(K)})$, 
and the $k$th \emph{menu element}, $B^{(k)}$,
 specifies allocation probabilities on bundles,
 $\alpha^{(k)}: 2^M\rightarrow [0,1]$, and a {\em price}, $\beta^{(k)}\in \mathbb{R}$.
Here, we allow randomization, where  $\alpha^{(k)}(S)\in[0,1]$ denotes the
  probability that bundle $S\in 2^M$ is assigned to the bidder in menu element $k$. 
  %
   %
  %
We refer to the menu $B$ as corresponding to a {\em menu-based representation 
of an auction.} The bidder with bid $b$ is assigned the element from menu $B$ that maximizes their utility according to the reported valuation: $k^*\in \arg\max_k \sum_{S\in 2^M}\alpha^{(k)}(S)b(S)-\beta^{(k)}$. We denote this optimal element by $(\alpha^*(b), \beta^*(b))$. 
The use of menu-based representations for auction design 
is without loss of generality and DSIC~\cite{hammond1979straightforward}.
%
The optimal auction design problem is to find a menu-based representation that 
maximizes  expected revenue, i.e., $\mathbb{E}_{v\sim F}[ \beta^{*}(v)]$. Deep menu-based methods~\cite{dutting2024optimal,shen2019automated} in the differentiable economics literature~\cite{zheng2022ai,finocchiaro2021bridging,wang2023deep,ivanov2024principal,zhang2024position,hossain2024multi,rahme2020auction,ivanov2022optimal,curry2022differentiable,duan2023scalable} learn to generate such menus by neural networks.


\textbf{Diffusion Models and Continuous Normalizing Flow}. Diffusion models have  emerged as a powerful class of generative AI methods, spurring notable advances in a wide range of tasks such as image generation~\cite{rombach2022high,esser2024scaling}, video generation~\cite{ho2022video,ceylan2023pix2video,ho2022imagen}, molecular design~\cite{gruver2024protein}, text generation~\cite{lou2024discrete}, and multi-agent learning~\cite{wang2024diffusion}. At their core, these models perform a {\em forward noising process} in which noise is incrementally added to training data over multiple steps, gradually corrupting the original sample.
%
A {\em reverse diffusion process} is then learned to iteratively remove noise, thereby reconstructing data from near-random initial states. In our setting, instead of reconstructing data, we extend the diffusion process to develop a tractable and differentiable method that optimizes a high-dimensional distribution.

In particular, {\em score-based diffusion models} enjoy strong mathematical and physical underpinnings. The forward noising process is an {\em Itô stochastic differential equation} (SDE),
\begin{align}
    d\vx = \vf(\vx,t)dt + h(t)d\vw,
\end{align}
where $\vx(t) \in\mathbb{R}^\ell$ is the {\em state} at time $t$, for some $\ell\in \mathbb{Z}_{>0}$, $\vf(\cdot,t):\mathbb{R}^\ell\rightarrow\mathbb{R}^\ell$ is the {\em drift coefficient}, $h(\cdot):\mathbb{R}\rightarrow\mathbb{R}$ is the {\em diffusion coefficient}, and $\vw$ is the {\em standard Wiener process} (Brownian motion). Different forward processes are designed by specifying functional forms for $\vf(\cdot,t)$ and $h(\cdot)$. The generation of data is then based on the reverse process, which is  a diffusion process  given by the {\em reverse-time SDE}~\cite{anderson1982reverse},
\begin{align}
    d\vx = [\vf(\vx,t)-h(t)^2\nabla_\vx\log q_t(\vx)]dt + h(t)d\bar{\vw},\label{equ:r-sde}
\end{align}
where $dt$ is an infinitesimal negative timestep,  $\bar{\vw}$ is the {\em standard Brownian motion with reversed time flow},  and {\em $q_t(\vx)$ is the distribution of state  $\vx(t)$
at time $t$}.
The principal task in diffusion models is to learn the {\em score function}, $\nabla_\vx\log q_t(\vx)$, which has been effectively achieved using neural networks in recent work. This
enables solving the reverse-time SDE and  generating new data samples.  Notably, in the diffusion model  (and more broadly, generative AI) literature, $q_0(\vx)$ is typically a known target distribution over data samples from a pre-training dateset.

The reverse-time SDE (Eq.~\ref{equ:r-sde}) can be mathematically intricate, motivating the study of an equivalent, \emph{deterministic reverse process} modeled by an ordinary differential equation (ODE),
\begin{align}
    d\vx = [\vf(\vx,t)-\frac{1}{2}h(t)^2\nabla_\vx\log q_t(\vx)]dt,\label{equ:r-ode}
\end{align}
which  preserves the same marginal probability densities $\{q_t(\vx)\}_{t=0}^T$ as the SDE in Eq.~\ref{equ:r-sde}~\cite{song2021score}.
%
%
Eq.~\ref{equ:r-ode} also highlights the connection between diffusion models and \emph{continuous normalizing flow}: each of them learns to transform and manipulate distributions by an ODE. Intuitively, a {\em continuous normalizing flow} transports an input $\vx_0\in \mathbb{R}^\ell$ to $\vx_t=\phi(t, \vx_0)$ at timestep $t\in[0,T]$.
Here, $\phi(t, \cdot):\mathbb{R}^\ell\rightarrow\mathbb{R}^\ell$ is the  \emph{flow}, and is governed by the ODE,
\begin{align}
    \frac{d}{dt}\vx_t = \varphi\left(t, \vx_t\right),\label{equ:f-ode}
\end{align}
where the vector field $\varphi: [0,T]\times \mathbb{R}^\ell\rightarrow \mathbb{R}^\ell$ specifies the  rate of
change of the state $\vx$.
Continuous normalizing flow \citep{chen2018neural} suggests to represent vector field $\varphi$ with a neural network. The flow $\phi$ transforms an initial distribution $p_0(\vx)$ to a final distribution $p_T(\vx)$ an time $T$.


\textbf{Rectified Flow}. A bottleneck that restricts the use of continuous normalizing flow in large-scale problems is that the ODE (Eq.~\ref{equ:f-ode})
is hard to solve when the vector field $\varphi$ is complex.  The {\em rectified flow}~\cite{liu2022flow} addresses this by encouraging the flow to follow the linear path:
\begin{align}
    \min_\varphi \int_0^T \mathbb{E}_{\vx_0\sim p_0(\vx),\vx_T\sim p_T(\vx)}\left[\|(\vx_T-\vx_0)-\varphi(t, \vx_t)\|^2\right]dt, \ \ \ \vx_t = t\vx_T + (1-t)\vx_0.\label{equ:rf}
\end{align}

Here, the target distribution $p_T(\vx)$ (from which $\vx_T$ are sampled) and the initial distribution $p_0(\vx)$ (from which $\vx_0$ are sampled) are known. $\vx_t,t\in[0,T]$ is the interpolated point between $\vx_T$ and $\vx_0$, and the rectified flow encourages the vector field to align as closely as possible with the straight line $\vx_T-\vx_0$.

As discussed in the introduction, the application of diffusion models or continuous normalizing flow in generative AI tasks relies on access to a known target distribution $p_T(\vx)$, but in our optimal CA design task, $p_T(\vx)$ is unknown and needs to be optimized.


%
\section{The Flow-Based Combinatorial Auction Menu Network}

As discussed in the introduction, the major challenge in learning menus for CAs is 
to provide an expressive representation of distributions over bundles
to associate with each menu element while retaining efficiency, so that
the exponential number of possible bundles does not become a bottleneck.
Moreover, training these representations adds another layer of difficulty: the menu must be not only concise but also easily differentiable to support training. 

\subsection{Menu representation}

Our key idea, following from score-based diffusion models and continuous normalizing flow, is to construct a concise and differentiable representation of a bundle distribution by modeling it through the solution of an ordinary differential equation (ODE). Specifically, the $k$th menu element generates its bundle distribution by the ODE,
\begin{equation}
    d\vs^{(k)}_t = \varphi^{(k)}(t,\vs^{(k)}_t) dt,\label{equ:de}
\end{equation}
for a suitable choice of vector field $\varphi^{(k)}$.
Here, we refer to $\vs^{(k)}_t\in \mathbb{R}^m$ as the \emph{bundle variable at time $t$}, where $m$ is the number of items. At time $T$, we require that a bundle variable $\vs^{(k)}_T$ represents a meaningful bundle, so that all entries are 0s or 1s,
and we adopt $\alpha^{(k)}_T(\vs^{(k)}_T)$ to denote the corresponding allocation probability. 
For simplicity, we omit the superscript $(k)$ when this is clear from the context.

By the  Liouville equation~\cite{liouville1838note}, the probability density at $T$ derived from Eq.~\ref{equ:de} satisfies:
\begin{align}
    \log \alpha_T(\vs_T) = \log \alpha_0(\vs_0) - \int_0^T \nabla\cdot \varphi(t, \vs_t) dt,\label{equ:liouville}
\end{align}
where $\alpha_0(\vs_0)$ denotes the initial distribution at time $0$, on initial bundle variables $\vs_0$, and $\nabla\cdot \varphi(t, \vs_t)$ is the divergence of $\varphi$.   Eq.~\ref{equ:liouville} is applicable to any $\vs_0$, and a bundle variable $\vs_t$ is generated from $\vs_0$ by $\vs_t(\vs_0)=\vs_0+\int_0^t \varphi(\tau,\vs_\tau) d\tau$. For clarity, we omit the explicit dependence on $\vs_0$ and simply write $\vs_t$.

\textbf{Training scheme}. Both the vector field $\varphi$ and the initial distribution $\alpha_0$ can influence the final distribution $\alpha_T$. 
Our method proceeds in two stages, involving the training of  each of
these two components in turn: 

$\quad$ \emph{(1) Flow Initialization.} We  fix  the initial distribution $\alpha_0$ and train the vector field $\varphi(t, \cdot)$ so that the final distribution, $\alpha_T$,
provides a reasonable coverage over bundles. 

$\quad$ \emph{(2) Menu Optimization.} We  fix the vector field from Stage 1, and backpropagate the revenue-maximizing loss through the flow to update the initial distribution $\alpha_0^{(k)}$ 
for each menu element $k$.


$\varphi$ and $\alpha_0(\vs_0)$ play a crucial role in maintaining a concise and easily differentiable representation and ensuring efficient training. We next propose specific functional forms for these two components that meet these criteria.

\textbf{Vector field}. We adopt the following functional form for the vector field,
\begin{align}
    \varphi(t, \vs_t;\xi,\theta) = \eta(t;\xi) Q(\vs_0;\theta) \vs_t,\label{equ:varphi}
\end{align}
where $Q: \mathbb{R}^{m}\rightarrow\mathbb{R}^{m\times m}$, written as a function of $\vs_0$, and the scalar factor $\eta:\mathbb{R}\rightarrow \mathbb{R}$, written as a function of  the ODE time $t\in[0,T]$, are neural networks with learnable parameters $\theta$ and $\xi$, respectively. 
We  omit dependence on $\theta$ and $\xi$ when the context is clear. This formulation's advantage becomes apparent when we consider its divergence:
\begin{align}
    \nabla\cdot \varphi(t,\vs_t) &= \sum_{i=1}^m \frac{\partial \varphi_i}{\partial s_{t,i}}
    = \sum_{i=1}^m \frac{\partial}{\partial s_{t,i}} \eta(t) Q_i(\vs_0) \vs_t = \sum_{i=1}^m  \eta(t) Q_{ii}(\vs_0) \\
    & = \eta(t)\Tr[Q(\vs_0)].
\end{align}

Here, $\varphi_i$ and $s_{t,i}$ are the $i$th element of $\varphi$ and $\vs_t$, respectively, $Q_i$ is the $i$th row of $Q$, and $Q_{ii}$ is the $i$th diagonal element of $Q$. Thus, the probability density at $T$ becomes
\begin{align}
    \log \alpha_T(\vs_T) = \log \alpha_0(\vs_0) - \Tr[Q(\vs_0)]\int_0^T \eta(t) dt.\label{equ:likelihood}
\end{align}

The integral in Eq.~\ref{equ:likelihood} 
is tractable as it only involves a scalar function, instead of bundle variables.
We can efficiently estimate this integral by time discretization. 

\textbf{Initial distribution}. In Stage 1, we use a mixture-of-Gaussian distribution for
the initial distribution $\alpha_0(\vs_0)$ on bundle variables $\vs_0$, with
\begin{align}
    \vs_0 \sim\sum_{d=1}^D w_d \mathcal{N}(\bm\mu_d, \sigma_d^2\mI_m),\label{equ:init_dist}
\end{align}
where, for $D$ components,
$\bm\mu_d\in\mathbb{R}^m$, $\sigma_d\in\mathbb{R}_{>0}$, $\mI_m$ is the $m\times m$ identity matrix, and $w_d\geq 0$ are weights satisfying $\sum_{d=1}^D w_d=1$. In Stage 2, as discussed later, we ensure DSIC by adopting a mixture-of-Dirac distribution, which is practically implemented by setting a very small variance $\sigma_d$ in a mixture-of-Gaussian distribution.


\subsection{Stage 1: Flow initialization}

The aim of the first stage is to guarantee that the flow can transport any initial bundle variable $\vs_0$ to a feasible bundle $S\in 2^M$.  We use $\vs=(\mathbb{I}{\{i\in S\}})$ to denote the vectorization of set $S$, i.e., the $i$-th component of $\vs$ is 1 if item $i$ is in $S$ and 0 otherwise.

In practice, numerical issues make it challenging to exactly obtain an feasible bundle $\vs$; i.e., a bundle variable
with only 0s and 1s. To account for this, we allow a small region around $\vs$ to be approximated as $\vs$ by modeling the bundle as a Gaussian variable,
\begin{align}
    S_{\sigma_z} = \mathcal{N}(\vs, \sigma_z^2\mI_m).
\end{align}

We train the vector field networks using rectified flow (\citet{liu2022flow}, Eq.~\ref{equ:rf}). For this  stage, we fix the initial distribution $\alpha_0(\vs_0)$ to a mixture-of-Gaussian model $\alpha_0(\vs_0)=\sum_{d=1}^D w_d \mathcal{N}(\bm\mu_d, \sigma_d^2\mI_m)$ with $D$ components.
We  define
$\alpha_T(\vs_T)$ as a uniform mixture-of-Gaussian model, with  components centered around each feasible bundle, and
$\alpha_T(\vs_T)=\frac{1}{2^m}\sum_{S\in 2^M} \mathcal{N}(\vs, \sigma_z^2\mI_m)=\frac{1}{2^m}\sum_{S\in 2^M}S_{\sigma_z}$.
This target distribution only applies in Stage 1, where it serves to encourage a balanced coverage of the final distribution over feasible bundles. In Stage 2, we have an optimization problem, and there is no longer a fixed target distribution.

We  follow the idea of  rectified flow, and define the {\em flow training loss} as
\begin{align}
    \mathcal{L}_{\textsc{Flow}}(\theta,\xi) =& \mathbb{E}_{(\vs_0,\vs_T)\sim (\alpha_0,\alpha_T), t\sim [0,T]} \left[\|(\vs_T-\vs_0)-\varphi(t, \vs_t; \theta,\xi)\|^2\right], \label{equ:flow_loss}\ \ \mbox{where}\\
    & \vs_t = t\cdot \vs_T + (1-t)\cdot \vs_0,\\
    & \varphi(t, \vs_t; \theta,\xi)=\eta(t;\xi)Q(\vs_0;\theta)\vs_t.
\end{align}

This loss is used to update the neural networks $Q$ and $\eta$ to encourage the vector field at interpolated points $\vs_t$ to point from $\vs_0$ to $\vs_T$.
%
The expectation in the flow training loss is taken over $(\alpha_0,\alpha_T)$, 
but directly sampling from $\alpha_T$ is intractable as it involves $2^m$ bundles.

Crucially, using a flow-based representation provides a workaround. We first draw $\vs_0\sim \alpha_0$, which is straightforward given that $\alpha_0$ comprises a manageable number of components ($D$). We then round $\vs_0$ to
the nearest feasible bundle, $\vs=\mathbb{I}(\vs_0\ge 0.5)\in\{0,1\}^m$,
and sample $\vs_T\sim \mathcal{N}(\vs, \sigma_z^2\mI_m)$. This approach underscores an advantage of deep learning. Although we cannot enumerate all possible bundles, the generalization ability of neural networks allows for learning the mapping from $\alpha_0$ to $\alpha_T$ given enough training samples.

\subsection{Stage 2: Menu optimization}\label{sec:method:opt}

In the second stage, we train the menu to seek to maximize the expected revenue for the auctioneer. For each menu element $k$,  the 
trainable parameters comprise the price $\beta^{(k)}$, as well as the parameters $w_d^{(k)}$ and $\bm\mu_d^{(k)}$ that define the initial distribution $\alpha^{(k)}_0$ on the bundle variable. 
The vector field $\varphi$ is  fixed in this stage and shared
among all menu elements.

Given a bidder with a value function $v$, the payment to the auctioneer is the price associated with the menu element that provides the highest utility to the bidder. 
Thus, computing the utility of each menu element is central to evaluating the revenue objective.
We always maintain a null menu element (zero allocation, zero price), which ensures 
  individual rationality (IR), so that the bidder has  non-negative expected
utility. 

Computing the expected 
utility corresponding to a menu element with bundle distribution $\alpha^{(k)}$ 
is intractable when done with a direct calculation,
because
\begin{equation}
    u^{(k)}(v) = \sum_{S\in 2^M} \alpha^{(k)}(S)v(S)
\end{equation}
requires enumerating $2^m$ bundles for a general valuation function.
However, with our flow-based representation, we can get the bundle allocation probabilities by applying the flow to the initial distribution. Specifically, we have
\begin{align}
    u^{(k)}(v) = \mathbb{E}_{\vs_0\sim \alpha^{(k)}_0, \vs=\mathbb{I}(\phi(T,\vs_0)\ge 0.5)} \left[v(\vs) \alpha^{(k)}_0(\vs_0)\exp\left(-\Tr[Q(\vs_0)]\int_0^T \eta(t) dt]\right)\right],\label{equ:u}
\end{align}
by applying the exponential operation to both sides of Eq.~\ref{equ:likelihood}. Here, $\phi(T,\vs_0)$ is the solution of the ODE solved by forward Euler,
\begin{align}
    \phi(T,\vs_0) = \vs_0 + Q(\vs_0)\int_0^T \eta(t)\vs_t dt,\label{equ:phi_T_s_0}
\end{align}
and $\vs=\mathbb{I}(\phi(T,\vs_0)\ge 0.5)$ is the rounded final bundle. Due to its simple form, a modern ODE solver can efficiently solve the ODE (Eq.~\ref{equ:phi_T_s_0}) in just a few steps. Therefore, the calculation of $u^{(k)}(v)$ becomes tractable when we make the initial distribution simple.

%

To ensure DSIC, we need to accurately calculate the expectation in Eq.~\ref{equ:u} to get the exact
utility to the bidder. We accomplish this by employing a mixture-of-Dirac distribution as the initial distribution, which has finite support. To implement this in practice, we set, for Stage 2 only, a very small variance to the Gaussian components in Eq.~\ref{equ:init_dist}, with $\sigma_d=1e\shortn 20$ for every component $d$. In this way, the utility can be obtained by enumerating over the finite support of the initial distribution:
\begin{align}
    u^{(k)}(v) = \sum_{d=1}^D \left[v(\vs(\bm\mu^{(k)}_d)) \alpha^{(k)}_0(\bm\mu^{(k)}_d)\exp\left(-\Tr[Q(\bm\mu^{(k)}_d)]\int_0^T \eta(t) dt]\right)\right],\label{equ:u_finite}
\end{align}
where $\vs(\bm\mu^{(k)}_d)=\mathbb{I}(\phi(T,\bm\mu^{(k)}_d)\ge 0.5)$. 
That is, the support of $\alpha^{(k)}_0$ 
consists, in effect, of the set of means, one for each component. 
It is worth noting that $D$ in Eq.~\ref{equ:u_finite} does not need to be the same $D$ as in Stage 1, and it could even vary across menu elements.

In this Stage 2, we fix the vector field $\varphi$ ($Q$ and $\eta$ networks) in Eq.~\ref{equ:u_finite} and update trainable parameters associated with the price and 
initial distribution $\alpha_0^{(k)}$ for each menu element $k$ 
during menu optimization, \ie, $\beta^{(k)}$, $\{w^{(k)}_d\}_{d=1}^{D}$, and $\{\bm\mu^{(k)}_d\}_{d=1}^{D}$. Therefore, given a set of bidder valuations $\mathcal{V}$,
the {\em revenue-maximization loss} is defined as
\begin{align}
    \mathcal{L}_{\textsc{Rev}}\left(\{\beta^{(k)}\}_{k=1}^K, \bigl\{ w_d^{(k)} \bigr\}_{\substack{d\in[D] \\ k\in[K]}},\bigl\{\bm\mu_d^{(k)} \bigr\}_{\substack{d\in[D] \\ k\in[K]}}\right) = -\frac{1}{|\mathcal{V}|}\sum_{v\in \mathcal{V}}\left[\sum_{k\in[K]}z^{(k)}(v)\beta^{(k)} \right],
\end{align}
where $z^{(k)}(v)$ is obtained by applying the differentiable SoftMax function to the utility of the bidder being allocated the $k$-th menu choice, i.e.,
\begin{align}
    z^{(k)}(v) = \mathsf{SoftMax}_k\left(\lambda_{\textsc{SoftMax}}\cdot u^{(1)}(v),\ldots,\lambda_{\textsc{SoftMax}}\cdot u^{(K)}(v)\right),\label{equ:softmax_in_loss}
\end{align}
where $\lambda_{\textsc{SoftMax}}$ is a scaling factor, and $u^{(k)}(v)$ is calculated by Eq.~\ref{equ:u_finite}.
When optimizing $\mathcal{L}_{\textsc{Rev}}$, the gradients with respect to $\beta^{(k)}$ are straightforward to compute. Moreover, although $Q$ remains fixed, gradients can still backpropagate through this network to update its input,
which is $\bm\mu^{(k)}_d$. Gradients also flow through $z^{(k)}$ back into $\alpha_0$, enabling updates to the mixture weights $w^{(k)}_d$. All these gradients are automatically handled by standard deep learning frameworks.


\subsection{Discussion}\label{sec:multi-bidder}


\textbf{DSIC}. The seminal work by \citet{hammond1979straightforward} establishes necessary and sufficient conditions for a strategyproof menu-based auction: (1) Self-bid independent: the menu is independent of the bidder's bid; (2) Agent-optimizing: the bidder is assigned the menu element that maximizes their utility. As we analyze here, our method satisfies these two properties.

In \name, all element prices, as well as bundle allocations, which depend on initial distributions and the vector field, are trained on values sampled from the distribution $F$, without using any information about the bidder's specific valuation. Therefore, menus learned by \name~are self-bid independent. 
As discussed in Sec.~\ref{sec:method:opt}, we require the initial distribution for each menu
element to have finite support, which means that the bundle distribution for each menu element 
can be reconstructed without any approximation error.
This guarantees exact utility calculation for every menu element. Moreover, unlike the SoftMax in Eq.~\ref{equ:softmax_in_loss}, we use hard argmax at test time, thereby selecting the menu element with the highest utility to the bidder. In this way, \name\ is strictly agent-optimizing.

\textbf{Expressiveness}. In Stage 1, we initialize the vector field $\varphi$. After this stage, given appropriate initial distributions, the final distribution can in principle cover all $2^m$ bundles and is trained to seek to achieve this.
In Stage 2, since the initial distribution for a menu element has finite support of size $D$, the bundle distribution for a menu element is also limited to finite support of size $D$. 
What is crucial, though, is that we can learn which (up to) $D$ bundles are represented in the distribution
that corresponds to a menu element. In practice, we find that a bounded  $D$
that is much smaller than $2^m$ still gives very high expected revenue.


\textbf{Extension to multi-bidder settings}. By providing an expressive and concise 
representation of single-bidder menus for the CA setting, our method opens up the possibilities of developing a general DSIC multi-bidder CA mechanism. A principled approach is to 
adapt the idea of GemNet~\cite{wang2024gemnet}. 
First, we can learn a separate \name~menu for each bidder. The modification in the network architecture is that these menus should now also depend on other bidders' bids $\vb_{\shortn i}$. To achieve this, we can condition the vector field, specifically the $Q$ and $\sigma$ networks, on $\vb_{\shortn i}$ by concatenating them to the inputs. For the price of each menu element, we can model them as the output of a neural network whose input is $\vb_{\shortn i}$. During training, we can also introduce a compatibility loss in the same way as that used in GemNet. This loss penalizes any over-allocation of items in the selected agent-optimizing elements from individual menus.

The major challenge in adapting GemNet to the CA setting  
arises during the post-training stage of GemNet, which adjusts prices of menu elements so that there is provably never any over-allocation of items. For this, GemNet constructs a grid over the space of bidder values. On each grid point, GemNet formulates a mixed-integer linear program (MILP) to adjust prices to ensure that, the utility of the best  element that is compatible with the choices of others in the sense of not over-allocating items is larger than that of all other elements by a safety margin. These safety margins prevent an incompatible menu element from being selected in the regions between grid points. Although the concise \name~menu representation, in principle, enables this MILP to 
be directly adapted to the combinatorial setting and used to adjust \name~menus to obtain a DSIC CA, the main issue is that the space of bidder values exhibits exponential dimensionality in the CA setting, resulting in an excessively large grid. Reducing this complexity represents the crucial 
remaining step in future work to enable 
a general, DSIC, and multi-bidder CA mechanism.
\section{Visualization on a didactic example}\label{sec:viz}

We present an example to visualize the training process of our method. For this, we adopt the \texttt{Regions} environment from the CATS testbed and consider uniform distributions with five items. 

Fig.~\ref{fig:flow_init}, already presented in the paper, shows the learning effect of Stage 1. Specifically, we fix the initial distribution and update the vector field. For the flow network architecture, the $Q$ and $\sigma$ networks have three and two 128-dimensional, tanh-activated, fully-connection hidden layers, respectively. We use the Adam optimizer with a learning rate of $5e\shortn 3$ to train these networks with $20$K samples for $60$K iterations.

The x- and y-axes of Fig.~\ref{fig:flow_init} are the bundle variables for the 3rd and 4th items, respectively. $x=1$ means that the 3rd item is in the bundle, and $y=1$ means the 4th item is in the bundle. The blue lines represent the vector field, and the blue dots represent the final distribution. We can see that the vector field gradually learns to cover all possible bundles in this projected, two-item subspace, 
as indicated by points $(0,0)$, $(0,1)$, $(1,0)$, and $(1,1)$ in the plots.

Fig.~\ref{fig:flow_train} illustrates the dynamics of Stage 2. In this demonstrative example, the support size of initial distributions is set to 512, the menu size is 8, and $\lambda_{\textsc{SoftMax}}$ is 1. The figure is organized into four columns, each displaying changes in the bundle allocation (depicted by blue dots) and  price of a distinct menu element. Additionally, the total revenue at the corresponding training iteration (during test time) is  presented. Changes in the bundle distribution for a menu element result from updates to the initial distribution for the menu element. For the initial distributions, the positions of the green dots represent the means ($\bm\mu_d^{(k)}$) of the mixture-of-Dirac components, while their opacities indicate the weights ($w_d^{(k)}$) of these components. We observe that each menu element learns to allocate different bundles. For example, Menu Element 3 manages bundles $(0,1)$ and $(1,0)$, and sets a price of 88.245 after Train Iteration 1020, at which point the revenue reaches 64.31.

\section{Experiments}

\begin{table}[t]
    \caption{Revenue comparison on CATS across different environments, value distributions, and numbers of items. Note that \emph{when $m=10$, the menu size is large enough to accommodate all possible bundles in a menu.} \label{tab:exp_results}}
    \vspace{-0.5em}
    \centering
    \begin{tabular}{ccccccc}
        \toprule
        \textbf{Environment} & \textbf{Baseline} & $m=10$ & $m=50$ & $m=75$ & $m=100$ & $m=150$ \\
        \midrule
        \multirow{5}{*}{\makecell{CATS-Regions \\ \footnotesize Uniform Private Valuations}} 
        & \grandbundle & 162.57 & 316.27 & 321.10 & 317.00 & 314.93 \\
        & \bigbundle & \textbf{202.26} & 399.85 & 354.48 & 322.68 & 329.17 \\
        & \smallbundle & 202.16 & 322.85 & 318.33 & 326.76 & 334.20 \\
        & \bundle & 189.17 & 288.41 & 290.14 & 292.11 & 312.65  \\
        & \name & 196.19 & \textbf{555.05} & \textbf{454.96} & \textbf{417.83} & \textbf{385.68} \\
        \midrule
        \multirow{5}{*}{\makecell{CATS-Regions \\ \footnotesize Normal Private Valuations}} 
        & \grandbundle & 142.54 & 319.06 & 328.89 & 305.06 & 309.73 \\
        & \bigbundle & \textbf{181.93} & 459.97 & 405.36 & 306.60 & 312.21 \\
        & \smallbundle & 181.75 & 342.31 & 339.82 & 303.31 & 315.02 \\
        & \bundle & 167.16 & 270.23 & 300.55 & 270.23 & 291.83 \\
        & \name & 173.93 & \textbf{603.70} & \textbf{448.35} & \textbf{389.51} & \textbf{394.82} \\
        \midrule
        \multirow{5}{*}{\makecell{CATS-Arbitrary \\ \footnotesize Uniform Private Valuations}} 
        & \grandbundle & 175.09 & 329.62 & 340.40 & 343.66 & 345.98 \\
        & \bigbundle &\textbf{233.26} & 396.78 & 351.70 & 357.44 & 351.07 \\
        & \smallbundle & 222.74 & 353.20 & 355.73 & 364.33 & 360.41 \\
        & \bundle & 205.02 & 316.79 & 312.41 & 313.76 & 334.51 \\
        & \name & 211.55 & \textbf{560.71} & \textbf{467.79} & \textbf{434.77} & \textbf{420.75} \\
        \midrule
        \multirow{5}{*}{\makecell{CATS-Arbitrary \\ \footnotesize Normal Private Valuations}} 
        & \grandbundle & 186.59 & 354.86 & 345.88 & 329.99 & 336.45 \\
        & \bigbundle & \textbf{248.16} & 478.67 & 349.34 & 335.75 & 339.10 \\
        & \smallbundle & 248.13 & 376.31 & 352.80 & 343.58 & 348.39 \\
        & \bundle & 221.49 & 348.41 & 312.79 & 304.00 & 316.81 \\
        & \name & 235.80 & \textbf{646.37} & \textbf{490.03} & \textbf{428.54} & \textbf{394.37} \\
        \bottomrule
    \end{tabular}
    \vspace{-0.5em}
\end{table}


\subsection{Testbeds}

\begin{figure}
    \centering
    \includegraphics[width=\linewidth]{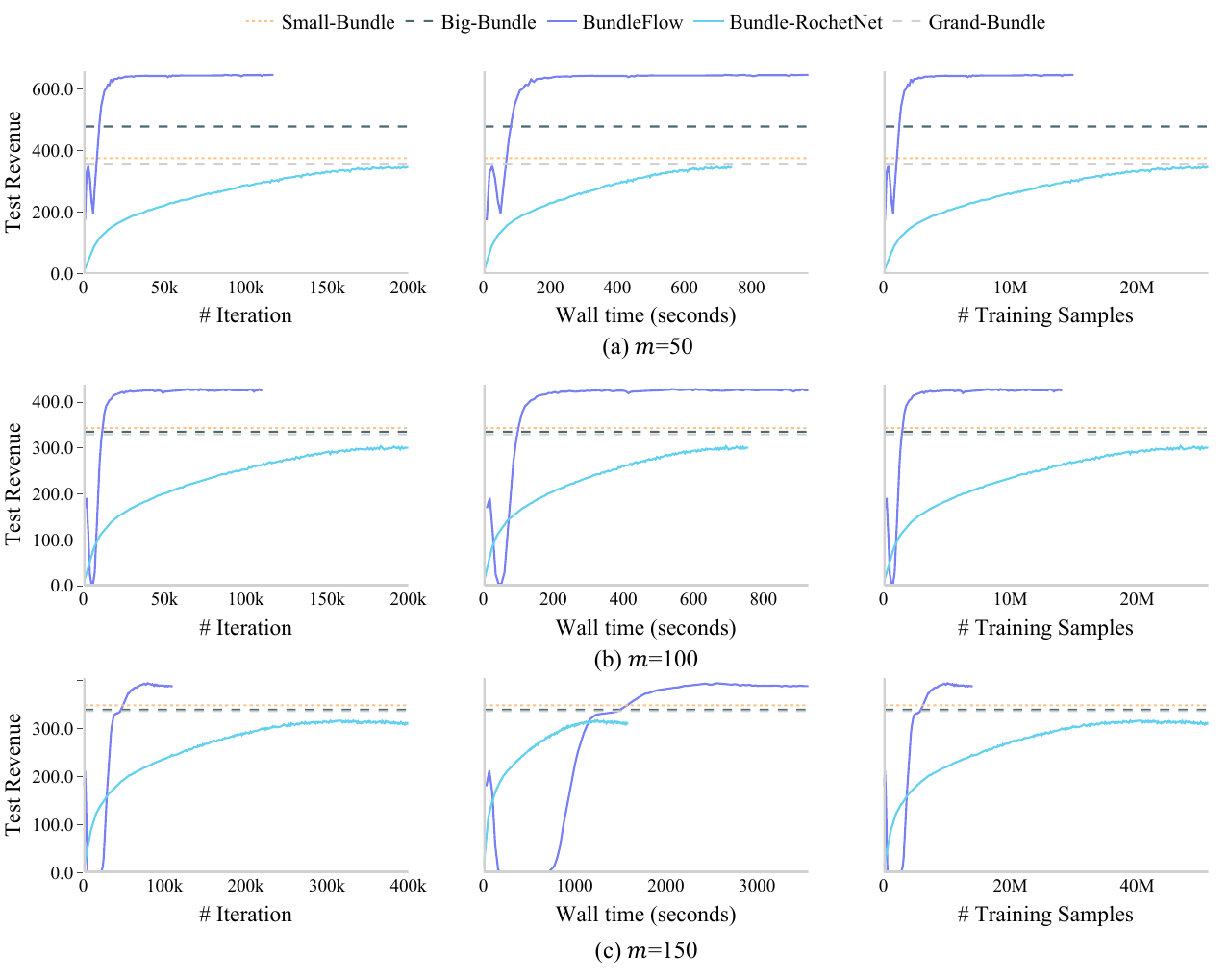}
    \caption{Learning curves of \name~and baselines on \texttt{CATS Arbitrary} with normal valuation distributions with different numbers of items. The three rows are results for 50, 100, and 150 items, respectively. The three columns show the changes of test revenue as a function of the number of training iterations, wall time in seconds, and the number of training samples, respectively. 
    \label{fig:lcm}}
\end{figure}

We evaluate our method on CATS \citep{leyton2000towards}, a standard benchmark that has been used in CA research since 2000. CATS has five different environments: \texttt{Regions}, \texttt{Arbitrary}, \texttt{Paths}, \texttt{Matching}, and \texttt{Scheduling}. These provide stylized representations of diverse real-world problems ranging from real estate and electronic components to transportation links and airline scheduling rights. Analysis of CA winner determination problem (WDP) complexity \citep{catshardness} revealed that \texttt{Matching}, \texttt{Scheduling}, and \texttt{Paths} are considerably easier to solve than \texttt{Regions} and \texttt{Arbitrary}. The WDP complexity was measured by CPLEX runs on problems with 256 goods and 1,000 
bids, reflecting the percentage of instances where the WDP is solved within a given time frame. Therefore, as in numerous previous works that have employed CATS as a benchmark \citep{gasse2019exact, hutter2009paramils, hutter2014algorithm, balcan2018learning, scavuzzo2022learning, wu2021learning, huang2023searching, song2020general, gupta2022lookback, balcan2021sample, zhang2022deep}, we focus our experiments on the challenging \texttt{Arbitrary} and  \texttt{Regions} environments. Each environment has two options for the value distribution: uniform and normal. We test both distributions with the default parameters provided by CATS.

In CATS, the valuation functions of bidders are recorded as bundle-bid pairs in an output file, with bundles from the same bidder identified by appending a dummy item tagged with a unique identifier. In effect, the bundle-bid pairs in an output file involving the same dummy item form an XOR representation of the bidder's valuation function. 
To obtain single-bidder valuations, we generate 100,000 such files and  extract valuation functions identified by a consistent dummy item. Of these, 95\% are used for training, with the remaining 5\% reserved for testing.


We evaluate our method with different numbers of items: 10, 50, 75, 100, and 150, across all environments and value distributions on CATS. When varying the number of items, we set the maximum XOR atoms per bid to 5 (the default value in CATS). We also experiment with a fixed number of items (50), increasing the maximum number of XOR atoms per valuation function in each environments by configuring the  \emph{maximum substitutable bids} argument.

\subsection{Baselines}

Work on DSIC deep auction learning can be broadly divided into two categories. (1) Methods like AMA~\cite{curry2022differentiable,duan2023scalable} and VVCA~\cite{duan2024scalable} explore a
restricted family of affine mechanisms. Consequently, they exhibit restricted expressiveness and achieve suboptimal revenue compared to methods in the second category~\cite{dutting2024optimal}.
(2) Menu-based methods like RochetNet~\cite{dutting2024optimal} learn an item-wise allocation for each menu element, and, as we have discussed, this requires an interpretation such as a product distribution for combinatorial valuations.
To test the flexibility of product distributions, we establish the first baseline as follows:

\begin{table}[t]
    \caption{Effect of the support size of the initial distribution ($D$) on \name~in different CATS environments. Revenue drops dramatically when $D$ decreases from 2 to 1. \label{tab:D}}
    \centering
    \begin{tabular}{ccccccc}
        \toprule
        \textbf{Environment} & \textbf{\# Items ($m$)} & $D=1$ & $D=2$ & $D=4$ & $D=8$ & $D=16$ \\
        \midrule
        \multirow{3}{*}{\makecell{CATS-Regions \\ \footnotesize Normal Private Valuations}} 
        & 50  & 278.82 & 589.50 & 596.67 & 603.70 & 595.14\\
        & 100 & 258.18 & 364.29 & 372.49 & 389.51 & 388.06 \\
        & 150 & 291.69 & 338.52 & 368.00 & 394.28 & 383.24 \\
        \midrule
        \multirow{3}{*}{\makecell{CATS-Arbitrary \\ \footnotesize Uniform Private Valuations}}
        & 50  & 287.98 & 557.06 & 563.70 & 560.71 & 561.51\\
        & 100 & 279.75 & 425.94 & 426.80 & 434.77 & 428.53  \\
        & 150 & 318.97 & 383.44 & 396.71 & 420.75 & 411.86\\
        \bottomrule
    \end{tabular}
\end{table}
\begin{table} [t]
    \caption{The Effect of increasing $D$ (the support size of initial distributions) under varying menu sizes $K$ (the number of elements in a menu). The default value of $K$ is 5000 when $m\le100$, and 20,000 otherwise.
    \label{tab:DJ} }
    \centering
    \begin{tabular}{CRCRCRCRCRCRCRCRCRCRCR}
        \toprule

        \multicolumn{4}{c}{Menu Size} &
        \multicolumn{4}{c}{$K/4$} & 
        \multicolumn{4}{c}{$K/2$} &
        \multicolumn{4}{c}{$K$} & 
        \multicolumn{4}{c}{$K*2$}\\

        \cmidrule(lr){1-4}
        \cmidrule(lr){5-8}
        \cmidrule(lr){9-12}
        \cmidrule(lr){13-16}
        \cmidrule(lr){17-20}
        \multicolumn{2}{c}{\textbf{Environments}} &
        \multicolumn{2}{l}{$m$} &
        \multicolumn{2}{c}{$D$=1} & 
        \multicolumn{2}{c}{$D$=2} &
        \multicolumn{2}{c}{$D$=1} & 
        \multicolumn{2}{c}{$D$=2} &
        \multicolumn{2}{c}{$D$=1} & 
        \multicolumn{2}{c}{$D$=2} &
        \multicolumn{2}{c}{$D$=1} & 
        \multicolumn{2}{c}{$D$=2} \\

        \cmidrule(lr){1-4}
        \cmidrule(lr){5-6}
        \cmidrule(lr){7-8}
        \cmidrule(lr){9-10}
        \cmidrule(lr){11-12}
        \cmidrule(lr){13-14}
        \cmidrule(lr){15-16}
        \cmidrule(lr){17-18}
        \cmidrule(lr){19-20}
        
        \multicolumn{2}{c}{\multirow{3}{*}{\makecell{CATS-Regions \\ \footnotesize Normal Private Valuations}}} & 
                               \multicolumn{2}{l}{50}  & \multicolumn{2}{l}{228.34} & \multicolumn{2}{l}{575.52} & \multicolumn{2}{l}{259.95} & \multicolumn{2}{l}{568.94} & \multicolumn{2}{l}{278.82} & \multicolumn{2}{l}{589.50} & \multicolumn{2}{l}{315.88}  & \multicolumn{2}{l}{602.98}\\
        \multicolumn{2}{c}{} & \multicolumn{2}{l}{100} & \multicolumn{2}{l}{213.26} & \multicolumn{2}{l}{324.04} & \multicolumn{2}{l}{226.11} & \multicolumn{2}{l}{345.51} & \multicolumn{2}{l}{258.18} & \multicolumn{2}{l}{364.29} & \multicolumn{2}{l}{270.65}  & \multicolumn{2}{l}{419.12}\\
        \multicolumn{2}{c}{} & \multicolumn{2}{l}{150} & \multicolumn{2}{l}{245.22} & \multicolumn{2}{l}{312.82} & \multicolumn{2}{l}{268.78} & \multicolumn{2}{l}{318.21} & \multicolumn{2}{l}{291.69} & \multicolumn{2}{l}{338.52} & \multicolumn{2}{l}{302.04}  & \multicolumn{2}{l}{385.08}\\
        
        \cmidrule(lr){1-4}
        \cmidrule(lr){5-6}
        \cmidrule(lr){7-8}
        \cmidrule(lr){9-10}
        \cmidrule(lr){11-12}
        \cmidrule(lr){13-14}
        \cmidrule(lr){15-16}
        \cmidrule(lr){17-18}
        \cmidrule(lr){19-20}
        
        \multicolumn{2}{c}{\multirow{3}{*}{\makecell{CATS-Arbitrary \\ \footnotesize Uniform Private Valuations}}} & 
                               \multicolumn{2}{l}{50}  & \multicolumn{2}{l}{238.96} & \multicolumn{2}{l}{546.50} & \multicolumn{2}{l}{261.61} & \multicolumn{2}{l}{551.78} & \multicolumn{2}{l}{287.98} & \multicolumn{2}{l}{557.06} & \multicolumn{2}{l}{303.53}  & \multicolumn{2}{l}{561.28}\\
        \multicolumn{2}{c}{} & \multicolumn{2}{l}{100} & \multicolumn{2}{l}{230.84} & \multicolumn{2}{l}{357.99} & \multicolumn{2}{l}{254.20} & \multicolumn{2}{l}{408.45} & \multicolumn{2}{l}{279.75} & \multicolumn{2}{l}{425.94} & \multicolumn{2}{l}{305.11}  & \multicolumn{2}{l}{438.27}\\
        \multicolumn{2}{c}{} & \multicolumn{2}{l}{150} & \multicolumn{2}{l}{297.22} & \multicolumn{2}{l}{355.07} & \multicolumn{2}{l}{303.08} & \multicolumn{2}{l}{361.01} & \multicolumn{2}{l}{318.97} & \multicolumn{2}{l}{383.44} & \multicolumn{2}{l}{343.19}  & \multicolumn{2}{l}{404.30}\\
        \bottomrule
    \end{tabular}
\end{table}

(1) \textbf{\bundle}: An adaption of RochetNet, interpreting its item-wise allocations as product distributions. Calculating bidder values remains intractable due to the need to enumerate all bundles. We address this by employing the \emph{Gumbel-SoftMax} technique~\cite{jang2017categorical}, which enables sampling from product distributions and backpropagation through the samples to update item-wise allocations. During training and testing, we use these samples to estimate bidder values. Empirically, item-wise allocations usually converge to binary vectors (0s or 1s) after training. In such scenarios, estimations of bidder values at test time are accurate, thereby ensuring that the mechanism is DSIC. If convergence to 0/1 vectors does not occur, DSIC is not exactly achieved in this baseline.


To better understand the performance of \name~and \bundle, we also fix the allocations in RochetNet menus and only learn the prices using the same gradient-based optimization method as in RochetNet. Specifically, we have the following additional baselines:

(2) \textbf{\bigbundle}: This baseline focuses on large bundles, including the grand bundle (that includes all items) and those nearest in size to the grand bundle. When the menu size prevents the inclusion of all bundles of a certain size, selection is random. 
For instance, with 3 items and a menu size of 3, the menu would include $[1,1,1]$ (the grand bundle) and a random subset of bundles containing $2$ items, such as $[1,1,0]$ and $[0,1,1]$.

(3) \textbf{\smallbundle}: Similar to \bigbundle, but it prioritizes minimal item allocation, along with the grand bundle; i.e., it begins with single-item bundles and expands as the menu size permits.

(4) \textbf{\grandbundle}: A simple baseline that sets a single price for the grand bundle. The price is determined through a grid search based on maximum training set revenue;  performance, as with all methods, is reported on the test set.
\begin{table}[t]
    \caption{Revenue comparison against baselines across different CATS environments and value distributions. The number of items is fixed at $m=50$, and we increase the valuation function size: $a = 10, 20, 30, 40$, and $50$ as maximum XOR atoms per valuation, corresponding to the {\em maxbid} parameter in CATS. \label{tab:exp_results_xor}}
    \centering
    \begin{tabular}{ccccccc}
        \toprule
        \textbf{Environment} & \textbf{Baseline} & $a=10$ & $a=20$ & $a=30$ & $a=40$ & $a=50$ \\
        \midrule
        \multirow{5}{*}{\makecell{CATS-Regions \\ \footnotesize Uniform Private Valuations}} 
        & \grandbundle & 314.46 & 309.11 & 316.04 & 320.61 & 314.35 \\
        & \bigbundle & 374.28 & 369.29 & 357.91 & 372.39 & 363.13 \\
        & \smallbundle & 328.95 & 328.92 & 323.44 & 333.00 & 326.58 \\
        & \bundle & 308.81 & 313.24 & 317.14 & 318.15 & 310.60 \\
        & \name & \textbf{533.42} & \textbf{528.46} & \textbf{528.26} & \textbf{539.92} & \textbf{537.36} \\
        \midrule
        \multirow{5}{*}{\makecell{CATS-Regions \\ \footnotesize Normal Private Valuations}} 
        & \grandbundle & 322.74 & 301.54 & 310.38 & 304.38 & 334.95 \\
        & \bigbundle & 434.32 & 375.51 & 404.74 & 381.05 & 430.24 \\
        & \smallbundle & 326.98 & 307.81 & 332.39 & 313.76 & 342.23 \\
        & \bundle & 301.47 & 297.34 & 317.10 & 302.04 & 327.32 \\
        & \name & \textbf{564.13} & \textbf{524.24} & \textbf{535.35} & \textbf{525.68} & \textbf{566.51} \\
        \midrule
        \multirow{5}{*}{\makecell{CATS-Arbitrary \\ \footnotesize Uniform Private Valuations}} 
        & \grandbundle & 331.21 & 334.43 & 330.33 & 351.01 & 336.68 \\
        & \bigbundle & 355.92 & 351.02 & 341.76 & 347.26 & 348.92 \\
        & \smallbundle & 352.34 & 354.58 & 346.36 & 354.87  & 353.58 \\
        & \bundle & 329.61 & 345.06 & 338.44 & 344.76 & 350.75 \\
        & \name & \textbf{565.54} & \textbf{583.60} & \textbf{568.01} & \textbf{579.01} & \textbf{581.39} \\
        \midrule
        \multirow{5}{*}{\makecell{CATS-Arbitrary \\ \footnotesize Normal Private Valuations}} 
        & \grandbundle & 381.51 & 335.44 & 323.52 & 320.28 & 358.83 \\
        & \bigbundle & 470.63 & 349.70 & 346.39 & 340.59 & 381.13 \\
        & \smallbundle & 402.23 & 341.79 & 337.44 & 332.90 & 368.26 \\
        & \bundle & 367.44 & 329.28 & 333.88 & 327.57 & 365.40 \\
        & \name & \textbf{664.94} & \textbf{564.90} & \textbf{552.77} & \textbf{548.22} & \textbf{615.93} \\
        \bottomrule
    \end{tabular}
\end{table}

\subsection{Setup}

We did not extensively fine-tune the flow model architecture, as the initial trial already yielded satisfactory results. This suggests that our formulation of the ODE, including its functional form (Eq.~\ref{equ:varphi}) and initial conditions (Eq.~\ref{equ:init_dist}), is well-suited to the needs of the CA settings, making the optimization of the flow model relatively straightforward.
Specifically, the $Q$ network comprises three 128-dimensional tanh-activated fully connected layers. When $m>100$, we increase the width of the last layer to 256. The $\sigma$ network is simpler and has two 128-dimensional tanh-activated fully connected layers. 

Two important hyper-parameters are $D$, the support size of the initial distribution, and $K$, the menu size. By default, $D$ is set to 8, a relatively small number. $K$ is 5000 when $m\le 100$ and is 20000 otherwise. This menu size setting is the same for our method, \smallbundle, \bigbundle, and \bundle.
Notably, the selected menu sizes are adequate to encompass all possible bundles for smaller numbers of items, such as $m=5$ or $10$. We will show the impact of different values of $D$ and $K$ in ablation studies. 

Menu optimization for \name~is conducted using the Adam optimizer with a learning rate of 0.3. $\lambda_{\textsc{SoftMax}}$ is increased from 0.001 to 0.2 over the course of training. For comprehensive details on our hyper-parameter settings, please refer to the codebase  provided with our submission. For the baselines, we fine-tuned their hyper-parameters so that they perform significantly better than the default RochetNet setting. The modifications are achieved by performing a grid search to obtain the optimum combination of $\lambda_{\textsc{SoftMax}}$ and learning rate that yields the best revenue and also guarantees convergence. Both \smallbundle~and \bigbundle~use a learning rate of $0.3$ and $\lambda_{\textsc{SoftMax}}$ of 2, while \bundle~uses a learning rate of $0.05$ and $\lambda_{\textsc{SoftMax}}$ of 20.


\subsection{Results}


\begin{figure}
    \centering
    \includegraphics[width=\linewidth]{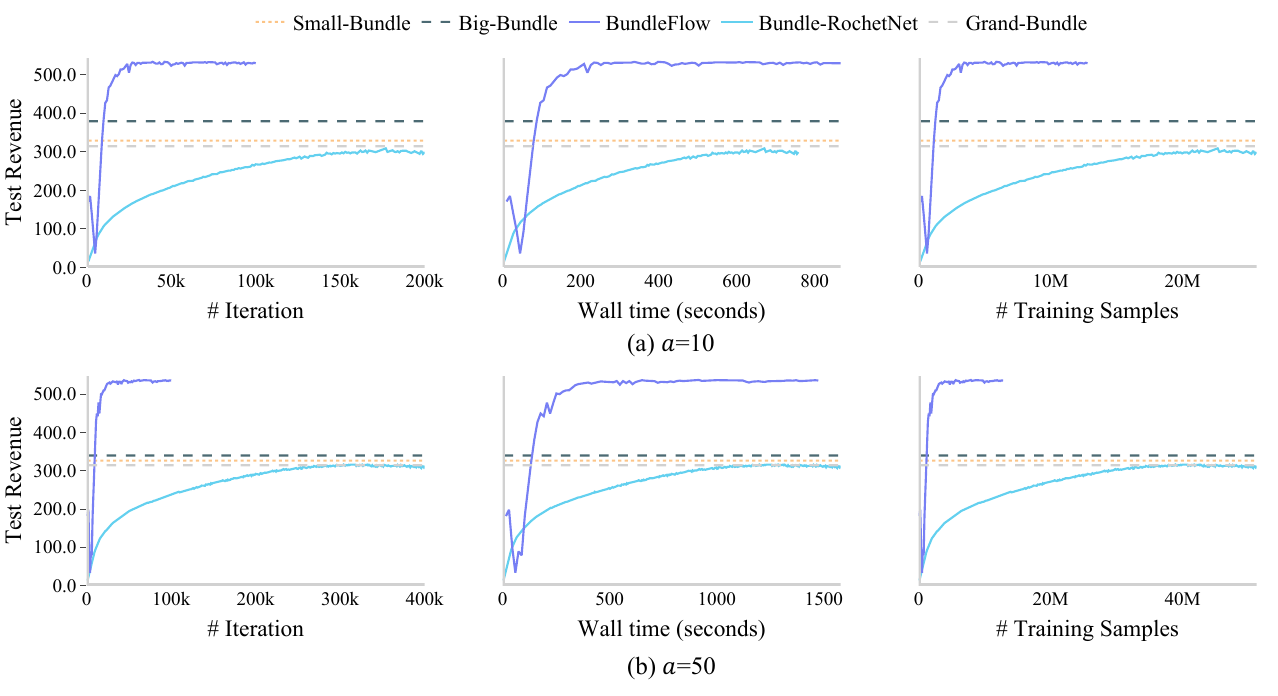}
    \caption{Learning curves of \name~and baselines on \texttt{CATS Regions} with uniform valuation distributions with different valuation function sizes. The two rows are results for $a=10$ and $50$ XOR atoms per valuation, respectively. The three columns show the changes of test revenue as a function of the number of training iterations, wall time in seconds, and the number of training samples, respectively.
    \label{fig:lca}}
\end{figure}

\textbf{Performance on CATS}. Table~\ref{tab:exp_results} shows the revenue of our method compared against baselines. Our method performs consistently and significantly better when the number of items is large ($m>10$), achieving revenues gains of 1.11$-$2.23$\times$. The learning curves in Fig.~\ref{fig:lcm} further illustrate this advantage. \bundle~is the baseline that also learns allocations in a menu. In CATS \texttt{Arbitrary} with normal distributions, \bundle~requires 200K iterations to converge for $m=50$ and $100$, and 250K iterations for $m=150$. In contrast, our method converges in 17K$-$21K iterations for $m=50$ and $100$ and 68K iterations for $m=150$. This translates to a 3.6$-$9.5$\times$ improvement in the number of training iterations required. Moreover, our method can also reduce training time in some settings. For example, when $m=50$, \name~reduces the training duration from about 700 seconds to 140 seconds, achieving a 5$\times$  improvement in training speed.


When the number of items is small ($m=10$), baselines with a menu size of 5000 can effectively cover all possible bundles. Although these settings are not the primary focus of our method, \name~can still achieve comparable revenue relative to the baselines, as shown in the first column of Table~\ref{tab:exp_results}.

Among the baselines, fixed allocation strategies (\smallbundle~and \bigbundle) outperform the full-fledged \bundle. A possible reason is that menu sizes of 5K or 20K remain negligible compared to the vast number of possible bundles like $2^{50}$ or $2^{150}$, but can pose a challenge for deep learning optimization. Therefore, the potential benefit of increased flexibility are over-weighed by the difficulty in optimization. This highlights an advantage of our method: while allowing better flexibility, it formulates an optimization problem that is more tractable, thereby enabling significantly improved performance.

\textbf{The support size of bundle distributions}. Table~\ref{tab:D} provides insights into why our method has superior performance. When we reduce the support size of initial distributions ($D$) to 1, each menu element deterministically assigns a single bundle to the bidder. This eliminates a key advantage of our method as it can no longer represent a distribution over bundles. Correspondingly, we observe a dramatic drop in revenue when $D$ is decreased from 2 to 1. For example, in the CATS \texttt{Regions} environment with normal distributions and 50 items, revenue declines sharply from 589.50 to 278.82. This trend remains consistent when we vary the menu size, as shown in Table~\ref{tab:DJ}. This pronounced performance gap between $D=1$ and $D=2$ highlights the importance of maintaining a randomized distribution over bundles in differentiable economics for CA settings---a capability uniquely enabled by our method.

Despite this loss of expressiveness when $D=1$, we find that our method can still outperform \bundle~in some settings. This is because, even when optimizing a deterministic bundle for each menu element, we are ``searching" directly within the bundle space and, in principle, can reach any possible bundle. By contrast, as we have discussed, product distributions can represent only a limited subset of possible bundles. Therefore, our method retains greater flexibility than product distributions even when $D=1$.

\textbf{Menu size and Valuation function size}. Table~\ref{tab:DJ} presents the performance of our method under varying menu sizes ($K$), specifically when they are halved or quartered, and when they are doubled. As discussed, a randomized distribution over bundles is crucial in each of these settings, as revenue drops sharply when $D$ decreases to 1. Furthermore, increasing the menu size tends to improve performance, although the gains are modest when the item count is relatively low. For example, in CATS \texttt{Regions} with normal distributions, when $m=100$, increasing the menu size from $K/4$ to $K*2$ leads to a $29.34\%$ performance boost (rising from 324.04 to 419.12), in contrast to a merely $4.77\%$ enhancement when $m=50$ (rising from 575.52 to 602.98). Increasing the maximum number of XOR atoms per valuation from $10$ to $50$ has minimal effects on the performance of both our method and the baselines, as evidenced in Table~\ref{tab:exp_results_xor} and Fig.~\ref{fig:lca}.

\section{Closing Remarks}

Our key contribution is the proposal of using an ODE to enable concise representation and efficient optimization for 
bundle distributions in menu-based CA deep learning, where these bundle distributions could in principle require an exponential representation size. We show that maintaining and effectively manipulating these  bundle distributions is important in achieving superior auction revenue. We also discuss possible directions for extending our method to a general multi-bidder CA design algorithm. We hope these ideas and findings can provide new inspiration and perspectives for research in combinatorial auctions.

\newpage
\bibliographystyle{ACM-Reference-Format}
\bibliography{sample-bibliography}

\newpage
\appendix

\end{document}